\relax
\documentstyle [art12]{article}
\def\be{\begin{equation}}
\def\bea{\begin{eqnarray}}
\def\ee{\end{equation}}
\def\eea{\end{eqnarray}}
\def\mp{\begin{minipage}}
\def\emp{\end{minipage}}
\begin{document}
\bibliographystyle {plain}

\thispagestyle{empty}

\begin{flushright}
{\sc OUTP}-96-33P\\
 hep-th/9606138
\end{flushright}
\vspace{.3cm}

\begin{center}
{\large\sc{ LOGARITHMIC OPERATORS AND DYNAMICAL EXTENTION
 OF THE SYMMETRY GROUP IN THE BOSONIC  $SU(2)_0$ AND SUSY $SU(2)_2$
WZNW MODELS.}  }\\[15mm]
{\sc J.-S. Caux, I. Kogan, A. Lewis and A. M. Tsvelik}\\[5mm]
{\it  Department of Physics, University of Oxford,\\Theoretical Physics,\\ 1 
Keble Road,\\  Oxford, OX1 3NP, UK} \\[15mm]

{\sc Abstract}

\begin{center}
\begin{minipage}{14cm}
We  study the operator product expansion in the bosonic $SU(2)_0$ 
and SUSY $SU(2)_2$ WZNW models. We find that these OPEs contain both 
logarithmic operators and new conserved currents, leading to an 
extension of the symmetry group.

\end{minipage}
\end{center}

\end{center}


\vfill
\newpage
\pagestyle{plain}
\setcounter{page}{1}

\renewcommand{\footnotesize}{\small}

\par
\section{Introduction}
 Dynamical generation of symmetries is quantum field theory 
is an intriguing subject. Such generation has  recently been observed in the 
Wess-Zumino-Novikov-Witten (WZNW) model on the U(r) group \cite{ckt}.
 The latter model is a critical two-dimensional theory arising form the
study of $2+1-d$ relativistic fermions interacting with a random
 non-Abelian gauge potential, and posesses a Kac-Moody current algebra. 
It turned out that  in the limit $r \rightarrow 0$ this theory  acquires an 
additional conserved current acting  as a generator of a continuous symmetry.
This additional symmetry is related to the spectral degeneracy and 
emergence of the so-called ``logarithmic operators'' \cite{g}.

 The  logarithmic behaviour of multipoint correlation 
functions in 
two-dimensional conformal field theory was first discussed  by
 Rozansky and  Saleur \cite{rs}.
 They found such behaviour for the  WZNW model on the supergroup
 GL(1,1).  Later the logarithms which once had been thought to 
be excluded altogether to preserve locality \cite{cf},  have
been found  in a multitude of other models such as the
  gravitationally  dressed CFTs \cite{bk},  $c=-2$ and more general
 $c_{p,1}$ models
\cite{g} \cite{f} \cite{k} \cite{gk}, the   critical
disordered models \cite{ckt} \cite{ms}, and might play a role in
the study of critical polymers and percolation  \cite{s} \cite{f}
\cite{k} \cite{w} and $2D$-magnetohydrodynamic turbulence \cite{rr}.
  They are also important for studying the problem of recoil
 in the theory of strings  and $D$-branes \cite{km} \cite{kmw}
\cite{p} \cite{emn} \cite{texas} as well as target-space symmetries 
 in string theory in general \cite{km}.

 It was suggested by Gurarie \cite{g} that logarithms appear when there is  
a degeneracy in the spectrum of conformal dimensions of the theory.
A consequence is the appearance of additional operators which, together
with ordinary primay ones, form the basis for the Jordan cell of the
Virasoro operator $L_0$. As we have already mentioned, in some cases 
spectral degeneracy has an additional effect producing extra symmetries. 

 In this paper we consider two types of critical models
 where such symmetries emerge. 
These are $SU(2)$ level $k = 0$ WZNW model and $SU(2)$ level 
 $k = 2$ supersymmetric 
WZNW model.  We approach these problems  studying  the operator product 
expansion (OPE) which follows from the solutions of 
Knizhnik-Zamolodchikov equations for multipoint correlation functions of 
the primary fields. We establish that at k = 0 (k = 2 for the SUSY theory) 
the emerging OPE contains additional current operators $K^a$ and $N^a$ 
(a = 1,2, 3). Therefore at these special values of k the current algebra 
enlarges to include 6 additional currents. We use these particular
 models as the simplest examples of two general classes - the 
 $SU(N)$ level $ k = 0$ WZNW model and $SU(N)$ level $k = N$ supersymmetric 
WZNW model. It can be shown that the only difference is that there are
  $2(N^2-1)$  new currents in the $SU(N)$ case instead of the $6$ currents
 we shall consider in this paper. The SUSY case  $SU(2)$ at level $k =2$
 was briefly discussed earlier in the  context of solitons in string 
theory \cite{km}

\section{Formulation of the theory}
 The generators for the Kac-Moody and Virasoro algebras for a 2-d
conformal field theory with zero central extension have the OPEs
\bea
T (z_1) T(z_2) &=& \frac{c/2}{z_{12}^4} + \frac{2}{z_{12}^2} T(z_2) 
+ \frac{1}{z_{12}} \partial
T(z_2) + ... \nonumber \\
T (z_1) J^a(z_2) &=& \frac{1}{z_{12}^2} J^a(z_1) + \frac{1}{z_{12}}
\partial J^a(z_2) + ... \nonumber \\
J^a (z_1) J^b(z_2) &=& \frac{f^{abc}}{z_{12}} J^c(z_2) + ...
\eea
We define our WZNW model via the Sugawara construction
\bea
T (z) = \frac{1}{2\kappa} : J^a (z) J^a (z): \label{sugawara}
\eea
and take the gauge group to be $SU(2)$.  
 Elementary calculations show that $\kappa = - \frac{c_v}{2} $ where
$c_v \delta^{cd} =  f^{abc} f^{abd}$, and that the central
charge of the Virasoro algebra vanishes. \\
 We assume the existence of primary fields tranforming as
\bea
T(z_1) \phi (z_2, \bar{z}_2) &=& \frac{\Delta}{z_{12}^2} \phi(z_2, \bar{z}_2) +
\frac{1}{z_{12}} \partial \phi (z_2, \bar{z}_2) + ... \nonumber \\
J^a (z_1) \phi (z_2, \bar{z}_2) &=& \frac{t^a}{z_{12}} \phi (z_2, \bar{z}_2) 
+ ... \label{exist}
\eea
In terms of the mode expansions $T(z) = \sum_n L_n/z^{n+2}$ and $J^a (z) =
\sum_n J^a_n /z^{n+1}$ the relation (\ref{sugawara}) reads
\bea
2 L_n = \sum_{m=-\infty}^{\infty} : J^a_m J^a_{n-m}: \nonumber 
\eea
with conventional normal ordering. \\
 Applying this to $g$ for $n = -1$, we find the existence of the null
vector
\bea
(J^a_{-1} t^a -  L_{-1}) \phi = 0
\eea
and see that the primary field has dimension $\Delta = 3/8$. 
  These, and the Ward Identities of conformal field
theory, directly lead to the Knizhnik-Zamolodchikov equations \cite{kz} 
for our case:
\bea
\left\{ \frac{c_v}{2} \partial_{z_i} + \sum_{i\neq j =1}^N \frac{t^a_i t^a_j}
{z_{ij}} \right\} <|\phi (z_1, \bar{z}_1)... \phi (z_N, \bar{z}_N) |> 
= 0 \label{kz}
\eea
 The full correlator can be built up starting from the chiral conformal
blocks by solving the monodromy problem to ensure locality \cite{df}.  This, 
contrary to what was anticipated a long time ago, can in some special
cases be done in a consistent way.  \\
 The conformal blocks for the level zero KZ equations were studied in
the context of a more general chain of thought in \cite{smirnov1}.  There,
an expression was obtained for the general $2n$-point conformal blocks.
The fields involved are taken to be from the fundamental spin-$1/2$ 
representation, and thus
carry an index $\epsilon$.  For a general conformal block of order $2n$,
the multiindex
$\epsilon_1,...,\epsilon_{2n}$ has to obey $\sum \epsilon_i = 0$ since
we are looking for singlet solutions of the equations.  There exists
then a partition of $B = \{1,...,2n\}$ into $T = \{i_k\}_{k=1}^n :
\epsilon_{i_k} = +$ and $T^{\prime} = \{j_k\}_{k=1}^n : \epsilon_{j_k}
= -$.  Distinct solutions are parametrized be the sets
$\gamma_1,...,\gamma_{n-1}$.
\begin{eqnarray}
f_{\gamma_1,...,\gamma_{n-1}}^{\epsilon_1,...,\epsilon_{2n}}
(\lambda_1,...,\lambda_{2n}) =&& \prod_{i<j} (\lambda_i -
\lambda_j)^{1/4} \times \nonumber \\ 
&&\prod_{k \in T, l \in T^{\prime}} (\lambda_k -
\lambda_l)^{-1} \det {\left| \int_{\gamma_p} \zeta_q (\tau | T |
T^{\prime}) \right|}_{(n-1)\times(n-1)} \label{2npt} 
\end{eqnarray}
where $\zeta_q$ are given by the following differentials on the hyper-elliptic
surface (HES) $w^2 = P(\tau) = \prod(\tau - \lambda_i)$:
\begin{eqnarray}
\zeta_q (\tau | T | T^{\prime}) &=& \frac{Q_q(\tau | T |
T^{\prime})}{\sqrt{P(\tau)}} \nonumber \\
Q_q (\tau) &=& \prod_{i \in T} (\tau - \lambda_i) {\left[ \frac{d}{d\tau}
\frac{\prod_{j \in T^{\prime}} (\tau - \lambda_j)}{\tau^{n - q}}
\right]}_0 + \nonumber \\ 
&&+ \prod_{i \in T^{\prime}} (\tau - \lambda_i) {\left[ \frac{d}{d\tau}
\frac{\prod_{j \in T} (\tau - \lambda_j)}{\tau^{n - q}}
\right]}_0 \label{zeta}
\end{eqnarray}
and where the $\gamma_i, i = 1,...,n-1$ contours are chosen to be
either of the set $\{a_i, b_i\}$:  $a_i$ surrounds the cut running
from $\lambda_{2i-1}$ and $\lambda_{2i}$, and $b_i$ starts from one
bank of the cut $\lambda_{2i-1}, \lambda_{2i}$, reaches the cut
$\lambda_{2g+1}, \lambda_{2g+2}$ and returns to the other bank
of $\lambda_{2i-1}, \lambda_{2i}$ by another sheet (the prescription
is given in \cite{smirnov2}).\\
 Given these, we can in principle study the full algebra of the theory.
We start this endeavour in the next section, and come to some surprises
quite soon along the way.

\section{Four-point correlation functions and Operator Product
Expansions}
 The integral representation (\ref{2npt}) allows us to calculate the
conformal blocks
\begin{eqnarray}
f^{\epsilon_1...\epsilon_4}(z_1,z_2,z_3,z_4) 
\sim \langle | V_{\epsilon_1} (z_1)...V_{\epsilon_4}(z_4)|\rangle
\end{eqnarray}
where $V$ is our primary chiral field, and $| \rangle$ is the $sl(2,C)$
-invariant vacuum from which we will build our physical Kac-Moody 
invariant correlator. \\
 As usual, since we want our correlation function to be invariant
under the Kac-Moody algebra, we decompose our correlator in terms of
amplitudes multiplied on the independent invariant tensors (basic 
singlets) of the representation.  The two independent contours   
$\gamma = a = [0, z]$ and $\gamma = b = [z, 1]$ then generate the 
independent solutions: 
\begin{eqnarray}
f_{\gamma}^{\epsilon_1...\epsilon_4}(z_1,z_2,z_3,z_4) =
\frac{1}{[z_{13}z_{24}]^{3/4}} f_{\gamma}^{\epsilon_1...\epsilon_4}(z)
\nonumber \\ 
f_{\gamma}^{\epsilon_1...\epsilon_4}(z) = \left[z(1-z)\right]^{1/4}
\sum_{A=1}^2 J_A F_{\gamma}^A (z) \nonumber \\
J_1 = \delta_{\epsilon_1 \epsilon_2}\delta_{\epsilon_3 \epsilon_4},
~~ J_2 = \delta_{\epsilon_1 \epsilon_4}\delta_{\epsilon_2 \epsilon_3}
\end{eqnarray}
where
\begin{eqnarray} 
F_a^{1}(z) &=&-\frac{1}{2} F(1/2, 3/2; 1; z) \nonumber \\
F_b^{1}(z) &=&\frac{\pi}{4}F(1/2, 3/2; 2; 1-z) \nonumber \\
&=&-\frac{1}{2} \ln z F(1/2, 3/2; 1; z) - \frac{1}{2} H_{0} (z)  \nonumber \\ 
F_a^{2}(z) &=&-\frac{1}{4}F(1/2, 3/2; 2; z) \nonumber \\
F_b^{2}(z) &=&\frac{\pi}{2} F(1/2, 3/2; 1; 1-z) \nonumber \\
&=& \frac{1}{z} -\frac{1}{4} \ln z F(1/2, 3/2; 2; z) -\frac{1}{4}
H_{1} (z)  \nonumber \\ 
H_{i} (z) &=& \sum_{n=0}^{\infty} z^n \frac{(1/2 )_n (3/2)_n}{n! (n+i)!}
\times  \left\{ \Psi (1/2 + n)  + \right. \nonumber \\ 
  && \left. + \Psi (3/2 + n) - \Psi (n+1) - \Psi (n+i+1) \right\} 
\end{eqnarray}
 The functions $F_b^{i}$ have logarithmic behavior near $z = 0$:
\begin{eqnarray}
F_a^1 (z) &=& -\frac{1}{2} \left( 1 + \frac{3}{4}z + O(z^2) \right)
\nonumber \\  
F_b^1 (z) &=& - \frac{1}{2} \ln z + \frac{\alpha}{2} + O(z \ln z) \nonumber \\ 
F_a^2 (z) &=& -\frac{1}{4} \left( 1 + \frac{3}{8}z + O(z^2) \right)
\nonumber \\
F_b^2 (z) &=& \frac{1}{z} -\frac{1}{4} \ln z + \frac{\alpha +1}{4} + O(z
\ln z)  
\end{eqnarray}
 where $\alpha = 4 \ln 2 - 2$. \\
 We wish for a little while to concentrate on the logarithmic
solutions.  
 To accomodate the logarithms in these we have to postulate
for the primary fields an OPE containing logarithmic operators:
\bea
V_{\epsilon_1} (z_1) V_{\epsilon_2} (z_2) = \frac{1}{z_{12}^{3/4}}
\left\{ I_{\epsilon_2 \epsilon_2^{\vee}} - 
z_{12} t^i_{\epsilon_2 \epsilon_2^{\vee}} \left[ D^i
(z_2) +
\ln z_{12} C^i (z_2) \right] + ... \right\} \label{OPEVV}
\eea
where $I$ is the unit matrix and $\epsilon^{\vee}$ is the weight 
conjugate to $\epsilon$.
Taking the four-point vertex function and performing the fusion, we
obtain
\bea
\langle | V_{\epsilon_1} (z_1) V_{\epsilon_2} (z_2) D^i (z_3) |
\rangle &=& t^i_{\epsilon_1 \epsilon_2^\vee} \left[\alpha
- \ln {\frac{z_{12}}{z_{13} z_{23}}} \right]
\frac{z_{12}^{1/4}}{z_{13}z_{23}} \nonumber \\
\langle | V_{\epsilon_1} (z_1) V_{\epsilon_2} (z_2) C^i (z_3) |
\rangle &=& -t^i_{\epsilon_1 \epsilon_2^\vee}
\frac{z_{12}^{1/4}}{z_{13}z_{23}}  
\eea
Fusing further, we get the two-point functions
\bea
\langle | D^i (z_1) D^j (z_2) | \rangle &=& - \left[ \alpha + 2 \ln
z_{24} \right] \frac{\delta^{ij}}{z_{12}^2} \nonumber \\
\langle | C^i (z_1) D^j (z_2) | \rangle &=& \frac{\delta^{ij}}{z_{12}^2}
\nonumber \\
\langle | C^i (z_1) C^j (z_2) | \rangle &=& 0
\eea
 We can now look at the action of the Virasoro algebra generators on
these structures.  We start from the commutation relation between the
Virasoro generators and a primary field:
\bea
\left[ L_k , B (z) \right] = \left( z^{k+1} \partial + (k+1)\Delta_B
z^k \right) B (z)
\eea
 Then, by mode-expanding our operators
\bea
C^i (z) = \sum \frac{C^i_n}{z^{n+1}} ~~~ D^i (z) = \sum \frac{D^i_n}{z^{n+1}}
\eea
we get the commutation relations
\bea
\left[ L_0, C^i_{-n} \right] &=& (n + 1) C^i_{-n} \nonumber \\
\left[ L_0, D^i_{-n} \right] &=& (n +1) D^i_{-n} + C^i_{-n} \nonumber \\
\left[ L_k, C^i_{-n-k} \right] &=& \left( n + 1 + \frac{3}{8} (k-1) \right) 
C^i_{-n}
\nonumber \\
\left[ L_k, D^i_{-n-k} \right] &=& \left( n + 1 + \frac{3}{8} (k-1) \right)
D^i_{-n} + C^i_{-n} \nonumber \\
\eea
for $k \geq 1$.  The operator $L_0$ thus has the operators $C$ and $D$ as a
basis for its Jordan cell \cite{g}.  These allow to express the 
states $|C^i, n>$ and
$|D^i, n>$ in terms of the descendants of $|C^i, 0>$ and $|D^i, 0>$.
The result for the first two levels is
\bea
|C^i, 1> &=& \frac{1}{2} L_{-1} |C^i, 0> \nonumber \\
|C^i, 2> &=& \left[ \frac{3}{8} L_{-2} - 
\frac{1}{48} L^2_{-1} \right] |C^i, 0>
\nonumber \\
|D^i, 1> &=& \frac{1}{2} L_{-1} |D^i, 0> \nonumber \\
|D^i, 2> &=& \left[ \frac{3}{8} L_{-2} - 
\frac{1}{48} L^2_{-1} \right] |D^i, 0>
 + \left[ - \frac{23}{24} L_{-2} + \frac{83}{144} L^2_{-1} \right] |C^i, 0>
\eea  
 We are interested in the ``full'' four-point
correlation functions, in which the anti-holomorphic dependence has
been reestablished:
\bea
G (z_i, \bar{z}_i) = \langle | g (z_1, \bar{z}_1) g^{\dagger} (z_2,
\bar{z}_2) g (z_3, \bar{z}_3) g^{\dagger} (z_4, \bar{z}_4) | \rangle
\nonumber
\eea
for the $SU(2)_0 $ WZNW model.
We write our primary field as $g_{\epsilon \bar{\epsilon}} (z, \bar{z})$
where the indices $\epsilon (\bar{\epsilon})$ correspond respectively
to the left (right) $SU(2)$ group:  i.e., in the above, $g(z, \bar{z})
= g_{\epsilon \bar{\epsilon}} (z, \bar{z}); g^{\dagger} (z, \bar{z}) =
g^{\dagger}_{\bar{\epsilon} \epsilon} (z, \bar{z})$.
 Because of the conformal Ward identities \cite{kz}, we can give the
following form to the four-point function:
\begin{eqnarray}
&&\langle |g_{\epsilon_1 \bar{\epsilon}_1}(z_1, \bar{z}_1)
g^{\dagger}_{\bar{\epsilon_2} \epsilon_2}(z_2, \bar{z}_2)g_{\epsilon_3
\bar{\epsilon}_3}(z_3,\bar{z}_3)g^{\dagger}_{\bar{\epsilon_4} \epsilon_4}(z_4,
\bar{z}_4) | \rangle = \nonumber \\
&& ~~~~~~~~~~~~~~~~~~~~~~~~~~~~~~~~~~~
|(z_1 - z_3)(z_2 - z_4)|^{-4\Delta}
G_{\epsilon_1 \bar{\epsilon}_1,...,\epsilon_4 \bar{\epsilon}_4}(z,
\bar{z})  \\
&&G_{\epsilon_1 \bar{\epsilon}_1,...,\epsilon_4 \bar{\epsilon}_4}(z,
\bar{z}) := \lim_{|z_{\infty}| \rightarrow \infty}
|z_{\infty}|^{4\Delta}  \langle |g_{\epsilon_1 \bar{\epsilon}_1}(0,0) 
g^{\dagger}_{\epsilon_2 \bar{\epsilon}_2}(z, \bar{z})g_{\epsilon_3
\bar{\epsilon}_3}(1,1)g^{\dagger}_{\epsilon_4 \bar{\epsilon}_4}(z_{\infty},
\bar{z}_{\infty}) | \rangle \nonumber
\label{4pt}
\end{eqnarray}
where $z = \frac{z_{12} z_{34}}{z_{13}z_{24}}$.  
We can here again develop our correlator in the same fashion as usual:
\bea
G_{\epsilon_1 \bar{\epsilon}_1... \epsilon_4 \bar{\epsilon}_4} (z,
\bar{z}) = |z(1-z)|^{1/2} \sum_{i,j=1,2} (I_i)_{\epsilon_1,...,\epsilon_4}
(\bar{I}_j)_{\bar{\epsilon}_1,...,\bar{\epsilon}_4} G^{ij} (z,\bar{z})
\nonumber \\ 
I_1 = \delta_{\epsilon_1 \epsilon_4} \delta_{\epsilon_2 \epsilon_3}  ~~~~~
I_2 = \delta_{\epsilon_1 \epsilon_2} \delta_{\epsilon_3 \epsilon_4}  
\eea
 To build the physical correlator, we have to solve the monodromy
problem to ensure locality.  The conformal blocks have to obey the 
crossing symmetry
\bea
G_{\epsilon_1 \bar{\epsilon}_1 \epsilon_2 \bar{\epsilon}_2 \epsilon_3
\bar{\epsilon}_3 \epsilon_4 \bar{\epsilon}_4} (z, \bar{z}) =
G_{\epsilon_1 \bar{\epsilon}_1 \epsilon_4 \bar{\epsilon}_4 \epsilon_3
\bar{\epsilon}_3 \epsilon_2 \bar{\epsilon}_2} (1-z, 1-\bar{z}) 
\eea
which leads to the construction
\bea
G^{ij}(z, \bar{z}) = \sum_{p,q = a,b} U^{pq} F_p^i (z)
F_q^j (\bar{z}) \nonumber \\
U^{ab} = U^{ba} = 1 ~~~~~ U^{aa} = U^{bb} =0
\eea
Single-valuedness around $z = 0$ and $z = 1$ has
respectively necessitated $U^{bb} = 0$ and $U^{aa} =0$, because we
cannot allow terms of the form $\ln z \ln {\bar{z}}$.  It is
interesting to note that this will lead directly to the absence of the
unit operator in the OPE of two primary fields $g$, and thus to the
vanishing of their two-point correlation function.  We have to switch
the sets of indices because our 
correlation function involves $g^{\dagger}$ at $z_2$ and $z_4$.  We thus
have the following behaviors near $z=0$:
\bea
G^{11} (z, \bar{z}) &=& \frac{1}{2} \ln |z| - \frac{\alpha}{2} + O(z
\ln z) \nonumber \\
G^{12} (z, \bar{z}) &=& - \frac{1}{2\bar{z}} + \frac{1}{4} \ln |z|
-\frac{\alpha}{4}  - \frac{1}{8} + O(z \ln z) \nonumber \\
G^{22} (z, \bar{z}) &=& - \frac{1}{4z} - \frac{1}{4\bar{z}} + \frac{1}{8} \ln
|z| - \frac{\alpha +1}{8} + O(z \ln z) \nonumber \\ 
\eea
The dominant terms can be accounted for with the OPE
\bea
g_{\epsilon_1 \bar{\epsilon}_1} (z_1, \bar{z}_1) g^{\dagger}_{\bar{\epsilon_2}
\epsilon_2} (z_2, \bar{z}_2) &=& |z_{12}|^{-3/2} \times \nonumber \\  
&\times& \left\{ z_{12} 
\delta_{\bar{\epsilon}_1 \bar{\epsilon}_2} t^i_{\epsilon_1
\epsilon_2} K^i (z_2) + O(z_{12}^2) \right. \nonumber \\
&&+  \bar{z}_{12} \delta_{\epsilon_1 \epsilon_2} 
\bar{t}^i_{\bar{\epsilon}_1 \bar{\epsilon}_2}
\bar{K}^i (\bar{z}_2) + O(\bar{z}_{12}^2)   \\
&&+ \left. |z_{12}|^2  t^i_{\epsilon_1
\epsilon_2} \bar{t}^j_{\bar{\epsilon}_1 \bar{\epsilon}_2}
\left[ D^{ij} (z_2, \bar{z}_2) + \ln |z_{12}| C^{ij} (z_2, \bar{z}_2)
\right] + ... \right\} \nonumber \label{ope}
\eea
In this, we use the set of
matrices $\{ t^a \}, a
= 1,2,3$ defined by 
\begin{equation}
t^a = \frac{\sigma^a}{2}, ~~~~~ [t^a, t^b] = i \epsilon^{abc}t^c
\end{equation}
 The term of $O(1/{|z|^2})$ is absent in $G^{11}$ because we were
forced to take $U^{bb} =0$.  As anticipated above, this prevents the
appearance of the unit operator in (\ref{ope}).  Consistency thus
leads to
\bea
\langle | g_{\epsilon_1 \bar{\epsilon}_1} (z_1, \bar{z}_1)
g^{\dagger}_{\bar{\epsilon_2} \epsilon_2} (z_2, \bar{z}_2) | \rangle = 0
\eea
 The appearance of the logarithm, as Gurarie has taught us, is due
to the presence of operators in the OPE (\ref{ope}) whose dimensions 
become degenerate.  There is an enlightening way to see how this comes
about, namely by taking the expression for the full correlator for finite
$k$ \cite{kz} and then going to the limit $k \rightarrow 0$.  The solution
reads
\bea
G (z_i, \bar{z}_i) = \frac{1}{|z_{13}z_{24}|^{4\Delta}} G (z, \bar{z}) 
&&~~~~~ \Delta = \frac{3}{4(2 +k)} \nonumber \\
G_{AB} (z, \bar{z}) &=& \sum_{p,q = 0,1} U_{pq} F_A^{(p)} (z) F_B^{(q)}
(\bar{z}) \nonumber \\
F_1^{(0)} (z) &=& x^{-\frac{3}{2(2+k)}} (1-x)^{\frac{1}{2(2+k)}}
F(\frac{1}{2+k}, -\frac{1}{2+k}; \frac{k}{2+k}; z) \nonumber \\
F_2^{(0)} (z) &=& \frac{1}{k} x^{\frac{1+2k}{2(2+k)}} (1-x)^{\frac{1}{2(2+k)}}
F(\frac{1+k}{2+k}, \frac{3+k}{2+k}; \frac{2+2k}{2+k}; z) \nonumber \\
F_1^{(1)} (z) &=& x^{\frac{1}{2(2+k)}} (1-x)^{\frac{1}{2(2+k)}}
F(\frac{1}{2+k}, \frac{3}{2+k}; \frac{4+k}{2+k}; z) \nonumber \\
F_2^{(1)} (z) &=& -2 x^{\frac{1}{2(2+k)}} (1-x)^{\frac{1}{2(2+k)}}
F(\frac{1}{2+k}, \frac{3}{2+k}; \frac{2}{2+k}; z) \nonumber \\
U_{10} &=& U_{01} = 0 ~~~ U_{11} = h U_{00} \nonumber \\
h &=& \frac{1}{4} \frac{\Gamma (\frac{1}{2+k}) \Gamma
(\frac{3}{2+k})}{\Gamma (\frac{1+k}{2+k}) \Gamma (\frac{-1+k}{2+k})}
\frac{\Gamma^2 (\frac{k}{2+k})}{\Gamma^2 (\frac{2}{2+k})}
\eea
 When we take the limit $k \rightarrow 0$, we have to deal with
gamma functions of vanishing arguments, and hypergeometric functions 
of vanishing $\gamma$ (third parameter).  The former is easily tackled
using the recursion relation
\bea
\Gamma (x+1) = x \Gamma (x) \nonumber 
\eea
giving, for example,
\bea
\Gamma (\frac{k}{2+k}) = \frac{2}{k} + 1 + \psi (1) + O(k) \nonumber 
\eea
where $\psi (x) = \frac{d}{dx} \ln \Gamma (x)$, whereas the latter
necessitates the use of standard identities like
\bea
\gamma (\gamma +1) F(\alpha, \beta; \gamma; z) = \gamma (\gamma + 1) F
(\alpha, \beta; \gamma +1; z) + \alpha \beta z F (\alpha +1, \beta +1;
\gamma +2; z) 
\eea
to yield, for example,
\bea
F (\frac{1}{2+k}, -\frac{1}{2+k}; \frac{k}{2+k}; z) =
- \frac{1}{2k} z F(1/2, 3/2; 2; z) + F(1/2, 1/2; 1; z) - \nonumber \\
-\frac{1}{2} \sum_{n=0}^{\infty} \frac{(1/2)_n (3/2)_n}{(2)_n} 
\frac{z^n}{n!} \left[ -\frac{1}{4n+2} + \frac{1}{2} - \frac{1}{2}
[\psi (n+2) - \psi (2)] \right] + O(k) \nonumber \\
\eea
 When writing down the full correlator as a power series in $k$, the
dominant term is of order $1/k^2$, but adds up to order $1/k$ with 
the identity
\bea
\lim_{k \rightarrow 0} \frac{z^k - 1}{k} = \ln z \nonumber
\eea
 Doing the expansion teaches us that there are operators in the OPE of 
$g, g^{\dagger}$ of dimensions $\Delta = \bar{\Delta} = \frac{2}{2+k}$ and 
$1$, i.e.
\bea
g(1) g^{\dagger} (2) = ... + |z_{12}|^{\frac{1}{2+k}} t^i t^j A^{ij} 
(2) + |z_{12}|^{\frac{1+2k}{2+k}} t^i t^j B^{ij} (2) + ... \nonumber 
\eea
which become degenerate as $k \rightarrow 0$, yielding the
logarithmic pair.
 Thus, the divergent term (of order $1/k$) in the expansion is just
our previous solution, as can be straightforwardly checked, and the
unit operator in (\ref{ope}) is suppressed as $k \rightarrow 0$ if 
we take a finite four-point function. \\
 
 From all this, we can read the
two-point functions between the newly 
introduced fields:
\bea
\langle | K^i (z_1) K^j (z_2) |\rangle &=&
\frac{\delta^{ij}}{z_{12}^2} \nonumber \\
\langle | \bar{K}^i (\bar{z}_1) \bar{K}^j (\bar{z}_2) |\rangle &=&
\frac{\delta^{ij}}{\bar{z}_{12}^2} \nonumber \\
\langle | C^{ik} (z_1, \bar{z}_1) D^{jl} (z_2, \bar{z}_2) |\rangle &=&
\frac{2 \delta^{ij} \delta^{kl}}{ |z_{12}|^4} \nonumber \\
\langle | D^{ik} (z_1, \bar{z}_1) D^{jl} (z_2, \bar{z}_2) |\rangle &=&
 - 2 \left[ \alpha + 2 \ln |z_{12}| \right]  \frac{\delta^{ij}
\delta^{kl}}{|z_{12}|^4} 
\label{2pt}
\eea
with all other correlators vanishing.\\
 We want to look now at the OPE's involving the Kac-Moody current.
The Kac-Moody current with zero central extension obeys the OPE
\bea
J^a (z_1) J^b (z_2) = \frac{i \epsilon^{abc}}{z_{12}} J^c (z_2)
\eea
The Ward Identity involving the Kac-Moody current then reads
\bea
\langle | J^a (z) J^{b_1}(\zeta_1) ... J^{b_n}(\zeta_n) g (1) g^{\dagger}(2)
 g(3) g^{\dagger}(4)|\rangle = \nonumber \\
=\sum_{k=1}^n \frac{i\epsilon^{ab_k c}}{z- \zeta_k} \langle | J^{b_1} (\zeta_1)
...J^{b_{k-1}} (\zeta_{k-1}) J^c (\zeta_k) ... J^{b_n} (\zeta_n) 
g (1) g^{\dagger}(2) g(3) g^{\dagger}(4) | \rangle + \nonumber \\
 + \sum_{j=1}^4 \frac{ t^a_{(j)}}{z - z_j} \langle| J...J 
g (1) g^{\dagger}(2) g(3)
g^{\dagger}(4)|\rangle 
\label{wi}
\eea
where the matrices $t^a_{(j)}$ act on the left indices of the fields $g$
(we have to remember that there is a sign difference between the
variations of $g$ and $g^{\dagger}$ under K-M transformations, which shows
up implicitely in the Ward Identity).  \\
 A straightforward calculation gives
\bea
\langle | J^a (z_1) K^b (z_2) K^c (z_3) |\rangle &=& \frac{i
\epsilon^{abc}}{z_{12} z_{13} z_{23}} \nonumber \\
\langle | J^a (z_1) J^b (z_2) K^c (z_3) K^d (z_4) | \rangle &=& 
\frac{\delta^{ab} \delta^{cd} z_{12} z_{34} - \delta^{ac} \delta^{bd}
z_{13} z_{24} + \delta^{ad} \delta^{bc} z_{14}
z_{23}}{z_{12}z_{13}z_{14}z_{23}z_{24}z_{34}} \nonumber \\
\langle | J^i (z_1) J^j (z_2) N^k (z_3) | \rangle &=& \frac{i
\epsilon^{ijk}}{z_{12}z_{13}z_{23}}
\label{3pt}
\eea
where we have introduced still the other field $N$ for consistency.
  From this we can
get the following set of OPEs:
\begin{eqnarray}
K^i (z_1) K^j (z_2) &=& \frac{\delta^{ij}}{z_{12}^2} +
\frac{i\epsilon^{ijk}}{z_{12}} N^k(z_2) + ... \nonumber \\
J^i (z_1) J^j (z_2) &=& \frac{i\epsilon^{ijk}}{z_{12}} J^k (z_2) +
... \nonumber \\
J^i (z_1) K^j (z_2) &=& \frac{i\epsilon^{ijk}}{z_{12}} K^k(z_2) +
... \nonumber \\
J^i (z_1) N^j (z_2) &=& \frac{\delta^{ij}}{z_{12}^2} +
\frac{i\epsilon^{ijk}}{z_{12}} N^k(z_2) + ... 
\end{eqnarray}
The other OPEs are still undetermined at this point:  we would need 
still the OPEs for $KN$ and $NN$, which entail knowledge of respectively
the six-point and eight-point correlation functions of $g$s.  We have
not worked out these functions at this stage.

\section{Supersymmetric WZW model}
The WZW model for $SU(2)$ at level $k$ is equivalent to the
bosonic sector of the supersymmetric WZW (SWZW) model at level $k+2$
\cite{{vkpr},{fuchs}}, and so we can
use the results of the previous section to find the conformal blocks
and OPE of the supersymmetric model at level 2. Unlike the bosonic 
$k=0$ model, the SWZW model at $k=2$ has a non-zero 
action, which can be written as (we use the notation of \cite{fuchs}):
\begin{equation}
S = \frac{k}{16\pi}\int dz d\bar{z} d\theta d\bar{\theta}
\left[ -Dg^s\bar{D}g^{s\dagger} + \int dt g^s
\frac{\partial g^{s\dagger}}{\partial t}(Dg^s\bar{D}g^{s\dagger}
+ \bar{D}g^sDg^{s\dagger} \right]
\label{action} \end{equation}
Here $g^s$ is a superfield in the fundamental representation
of $SU(2)$.
The super-Virasoro algebra is generated by the super-stress tensor
$T^s (z,\theta)$,
which has  the OPE 
\be
T^s (Z_1) T^s(Z_2) = \frac{\hat{c}}{4Z_{12}^3} +
\frac{3\theta_{12}}{2Z_{12}^2} T^s(Z_2) + \frac{1}{2Z_{12}} DT^s(Z_2)
+ \frac{\theta_{12}}{Z_{12}} \partial T^s(Z_2) + \cdots 
\ee
For the model (\ref{action}) with $k=2$, the superconformal charge
$\hat{c}=1$. There is also a super-Kac-Moody algebra, generated by
supercurrents $J^{sa}$, with  the the OPEs:
\bea
T^s (Z_1) J^{sa}(Z_2) &=& \frac{\theta_{12}}{2Z_{12}^2} J^{sa}(Z_2)
+ \frac{1}{2Z_{12}} DJ^{sa}(Z_2)
+ \frac{\theta_{12}}{Z_{12}} \partial J^{sa}(Z_2) + \cdots \nonumber \\
J^{sa} (Z_1) J^{sb}(Z_2) &=& \frac{\delta^{ab}}{Z_{12}}
+ \frac{\theta_{12}}{Z_{12}}f^{abc} J^{sc}(Z_2) + \cdots
\end{eqnarray}
In these formulas, $Z = (z,\theta)$, $\theta_{12} = \theta_1 -
\theta_2$ and $Z_{12}=z_1 - z_2 - \theta_1\theta_2$.
$D = \frac{\partial}{\partial\theta} + \theta\frac{\partial}{\partial z}$
and $\partial = \frac{\partial}{\partial z}$.
$T^s$ and $J^{sa}$ are superfields which contain
the bosonic stress tensor and currents:
\begin{eqnarray}
T^s(Z) = \Theta(z) + \theta T(z) \nonumber \\
J^{sa}(Z) = j^a(z) + \theta J^a(z)
\end{eqnarray}
We are interested in the four-point correlation function of $g^s$. 
It was shown in \cite{fuchs} that, as a consequence of the superconformal
Ward Identities, the four-point function in the SWZW model at 
level $k$ can be expressed in terms of the corresponding function
in the WZW model at level $k-2$. In the case of $k=2$, the result is: 
\begin{equation}
\langle |g^s(Z_1, \bar{Z}_1)
g^{s\dagger}(Z_2, \bar{Z}_2)g^s(Z_3,\bar{Z}_3)g^{s\dagger}(Z_4,
\bar{Z}_4) | \rangle = |Z_{13}Z_{24}|^{-4\Delta}
{\cal{G}}
(Z,\theta,\theta',\bar{Z},\bar{\theta},\bar{\theta}') 
\end{equation}
\begin{eqnarray}
{\cal{G}}(Z,\theta,\theta',\bar{Z},\bar{\theta},\bar{\theta}')
&=& |Z(1-Z)|^{1/2} \sum_{i,j=1,2} I_i
\bar{I}_j{\cal{G}}^{ij} (Z,\theta,\theta',\bar{Z},\bar{\theta},\bar{\theta}')
\nonumber \\ 
I_1 = \delta_{\epsilon_1 \epsilon_4} \delta_{\epsilon_2 \epsilon_3}  &&
I_2 = \delta_{\epsilon_1 \epsilon_2} \delta_{\epsilon_3 \epsilon_4}   \\
{\cal{G}}^{ij}(Z,\theta,\theta',\bar{Z},\bar{\theta},\bar{\theta}')
&=& [1 + {\cal{Q}}(1-Z)^{-1/2}\theta\theta']G^{ij}(Z,\bar{Z}) 
[1 + {\cal{Q}}^T(1-Z)^{-1/2}\theta\theta'] \nonumber  
\end{eqnarray}
Where
\begin{eqnarray}
&Z = \frac{Z_{12} Z_{34}}{Z_{13}Z_{24}} \nonumber \\
&\theta = \theta_{124} ~~~~~~~  \theta' = \theta_{123} \nonumber \\
&\theta_{ijk} = (Z_{ij}Z_{jk}Z_{ki})^{-1/2}(\theta_iZ_{jk} + \theta_jZ_{ki} +
\theta_kZ_{ij} +\theta_i\theta_j\theta_k) \nonumber \\ 
&{\cal{Q}} = \left( \begin{array}{cc} 
3/4 & 0 \\ 1/2 & -1 \end{array} \right) 
\end{eqnarray}
$G^{ij}(Z,\bar{Z})$ is the same function as
in (\ref{4pt}). The $\theta$-independent part of the four-point function
in the supersymmetric model is therefore identical to the four-point
function in the bosonic model. The OPE for the supersymmetric model 
is therefore essentially the same as (\ref{ope}):
\bea
g^s_{\epsilon_1 \bar{\epsilon}_1} (Z_1, \bar{Z}_1) 
g^{s\dagger}_{\bar{\epsilon_2}
\epsilon_2} (Z_2, \bar{Z}_2) &=& |Z_{12}|^{-3/2} \times \nonumber \\ 
&\times& \left\{ Z_{12} 
\delta_{\bar{\epsilon}_1 \bar{\epsilon}_2} t^i_{\epsilon_1
\epsilon_2} K^{si} (Z_2)
 + \bar{Z}_{12} \delta_{\epsilon_1 \epsilon_2} 
\bar{t}^i_{\bar{\epsilon}_1 \bar{\epsilon}_2}
\bar{K}^{si} (\bar{Z}_2) + \right.  \\
 &+&\left. |Z_{12}|^2  t^i_{\epsilon_1
\epsilon_2} \bar{t}^j_{\bar{\epsilon}_1 \bar{\epsilon}_2}
\left[ D^{sij} (Z_2, \bar{Z}_2) + \ln |Z_{12}| C^{sij} (Z_2, \bar{Z}_2)
\right] + \cdots \right\}
\label{sope} \nonumber
\eea
The non-zero two-point functions are:
\bea
\langle | K^{si} (Z_1) K^{sj} (Z_2) |\rangle &=&
\frac{\delta^{ij}}{Z_{12}^2} \nonumber \\
\langle | \bar{K}^{si} (\bar{Z}_1) \bar{K}^{sj} (\bar{Z}_2) |\rangle &=&
\frac{\delta^{ij}}{\bar{Z}_{12}^2} \nonumber \\
\langle | C^{sik} (Z_1, \bar{Z}_1) D^{sjl} (Z_2, \bar{Z}_2) |\rangle &=&
\frac{2 \delta^{ij} \delta^{kl}}{ |Z_{12}|^4} \nonumber \\
\langle | D^{sik} (Z_1, \bar{Z}_1) D^{sjl} (Z_2, \bar{Z}_2) |\rangle &=&
 - 2 \left[ \alpha + 2 \ln |Z_{12}| \right]  \frac{\delta^{ij}
\delta^{kl}}{|Z_{12}|^4} 
\label{s2pt}
\eea
The Ward Identity involving $J^{sa}$ is similar to (\ref{wi})
\bea
\langle | J^{sa} (Z_0) g^s (1) g^{s\dagger}(2) ... g^{s\dagger}(2n)|\rangle =
\sum_{j=1}^{2n} \frac{\theta_{0j} t^a_{(j)}}{Z_{0j}} 
\langle| g^s (1) g^{s\dagger}(2) ... g^{s\dagger}(2n)|\rangle 
\label{swi}
\eea
From (\ref{swi}) with $n=2$ and (\ref{sope}), we can calculate
the three-point function:
\begin{equation}
\langle | J^{sa} (Z_1) K^{sb} (Z_2) K^{sc} (Z_3) |\rangle = \frac{-i
\epsilon^{abc}\theta_{123}}{Z_{12}^{1/2} Z_{31}^{1/2} Z_{23}^{3/2}}
\label{s3pt}
\end{equation}
Of course (\ref{s2pt}) and (\ref{s3pt}) are just the supersymmetric 
generalisations of (\ref{2pt}) and (\ref{3pt}). To calculate
correlation functions involving two factors of the super-current 
$J^{sa}$, we have to take into account the central extension in the 
OPE
\begin{equation}
J^{sa} (Z_1) J^{sb}(Z_2) = \frac{\delta^{ab}}{Z_{12}}
+ \frac{\theta_{12}}{Z_{12}}i\epsilon^{abc} J^{sc}(Z_2) + ...
\label{JJ}
\end{equation}
This leads to the correlation function:
\begin{eqnarray}
\langle | J^{sa} (Z_1) J^{sb} (Z_2) K^{sc} (Z_3) K^{sd} (Z_4) |
\rangle &=&
\frac{\delta^{ab}\delta^{cd}}{Z_{12}Z_{34}^2} + \nonumber \\ 
+{Z_{12}^{-1/2}Z_{13}^{-1/2}Z_{14}^{-1/2}Z_{23}^{-1/2}
Z_{24}^{-1/2}Z_{34}^{-3/2}}
&\times& \left[
\delta^{ab} \delta^{cd} Z_{12}^{1/2}Z_{34}^{1/2} 
\theta_{134}\theta_{234} \right. \nonumber \\
&&~~+\delta^{ac} \delta^{bd}Z_{13}^{1/2} Z_{24}^{1/2}
\theta_{124}\theta_{234}
\nonumber \\ 
&&~~\left.- \delta^{ad} \delta^{bc} Z_{14}^{1/2} Z_{23}^{1/2}
\theta_{123}\theta_{234}\right]
\label{JJKK}
\end{eqnarray}
The first term has no counterpart in the bosonic theory as it
is a consequence of the extra term in (\ref{JJ}). The other three
terms are the supersymmetric equivalent of (\ref{3pt}).When we fuse
$K^{sc}(Z_3)$ and $K^{sd}(Z_4)$ in (\ref{JJKK}), the extra
term only contributes to the 2-point function 
$\langle | J^{sa}(Z_1)J^{sb}(Z_2) | \rangle$,
and so we find 
\begin{equation}
\langle | J^{si} (Z_1) J^{sj} (Z_2) N^{sk} (Z_3) | \rangle = \frac{-i
\epsilon^{ijk}\theta_{123}}{Z_{12}^{1/2}Z_{23}^{1/2}Z_{31}^{1/2}}
\end{equation}
From these correlation functions, we can extract the following OPEs:
\begin{eqnarray}
K^{si} (Z_1) K^{sj} (Z_2) &=& \frac{\delta^{ij}}{Z_{12}^2} -
\frac{i\epsilon^{ijk}\theta_{12}}{Z_{12}^2} N^{sk}(Z_2) + 
\frac{i\epsilon^{ijk}}{Z_{12}}DN^{sk} + ... \nonumber \\ 
J^{si}(Z_1) K^{sj}(Z_2) &=& 
\frac{i\epsilon^{ijk} \theta_{12}}{Z_{12}}K^{sb} + ... \nonumber \\
J^{si}(Z_1) N^{sj}(Z_2) &=&  \frac{\delta^{ij}}{Z_{12}}
+ \frac{\theta_{12}}{Z_{12}}i\epsilon^{ijk} N^{sk}(Z_2) + ...
\label{sKJN}
\end{eqnarray}
Comparing (\ref{sKJN}) and (\ref{JJ}), it can be seen that in the case
of the SWZW model we can consistently identify $N^{sa}(Z)$ with
$J^{sa}(Z)$. However, we cannot know if this is correct without 
knowing the OPE of $N^{sa}$ with itself, for which we would 
need to know six-point and eight-point correlation functions
in the SWZW model.
To rewrite these OPEs in terms of bosonic and fermionic 
componenent fields, we note that $J^{sa}$ and $N^{sa}$ have 
superconformal dimension $1/2$, and $K^{sa}$ has dimension $1$.
In other words, the bosonic components 
all have conformal dimension 1, but the fermionic component
of $K^s$ has dimension $3/2$ while the fermionic components
of $J^s$ and $N^s$ have dimension $1/2$. We can therefore 
decompose $J^s$, $K^s$ and $N^s$ as:  
\begin{eqnarray}
J^{sa}(Z) &=& \theta J^a(z) + j^a(z) \nonumber \\ 
K^{sa}(Z) &=& K^a(z) + \theta k^a(z) \nonumber \\
N^{sa}(Z) &=& \theta N^a (z) + n^a(z)
\end{eqnarray}
The OPE for the component fields is then:
\begin{eqnarray}
J^a(z_1) J^b(z_2) &=& \frac{\delta^{ab}}{z_{12}^2} + 
\frac{i\epsilon^{abc}}{z_{12}} J^c(z_2) + \cdots \\
K^a(z_1) K^b(z_2) &=& \frac{\delta^{ab}}{z_{12}^2} + 
\frac{i\epsilon^{abc}}{z_{12}} N^c(z_2) + \cdots \\
J^a(z_1) K^b(z_2) &=& \frac{i\epsilon^{abc}}{z_{12}} K^c(z_2) + \cdots \\
J^a(z_1) N^b(z_2) &=& \frac{\delta^{ab}}{z_{12}^2} + 
\frac{i\epsilon^{abc}}{z_{12}} N^c(z_2) + \cdots \\
j^a(z_1) j^b(z_2) &=& \frac{\delta^{ab}}{z_{12}} + \cdots \\
k^a(z_1) k^b(z_2) &=& \frac{2\delta^{ab}}{z_{12}^3} + \cdots \\
j^a(z_1) k^b(z_2) &=& \frac{-i\epsilon^{abc}}{z_{12}} J^c(z_2) + \cdots \\
j^a(z_1) n^b(z_2) &=& \frac{\delta^{ab}}{z_{12}} + \cdots \\
j^a(z_1) J^b(z_2) &=& \frac{-i\epsilon^{abc}}{z_{12}} j^c(z_2) + \cdots \\
j^a(z_1) K^b(z_2) &=& \cdots \\
j^a(z_1) N^b(z_2) &=& \frac{-i\epsilon^{abc}}{z_{12}} n^c(z_2) + \cdots \\
J^a(z_1) k^b(z_2) &=& \frac{i\epsilon^{abc}}{z_{12}} k^c(z_2) + \cdots \\
J^a(z_1) n^b(z_2) &=& \frac{i\epsilon^{abc}}{z_{12}} n^c(z_2)  + \cdots \\
K^a(z_1) k^b(z_2) &=& \frac{i\epsilon^{abc}}{z_{12}^2} n^c(z_2) + 
\frac{i\epsilon^{abc}}{z_{12}} \partial n^c(z_2) + \cdots
\end{eqnarray}
If we identify $J^{sa}=N^{sa}$, the bosonic components reduce to an 
$SO(4)_2$ current algebra:
\begin{eqnarray}
J^a(z_1) J^b(z_2) &=& \frac{\delta^{ab}}{z_{12}^2} + 
\frac{i\epsilon^{abc}}{z_{12}} J^c(z_2) + \cdots \nonumber \\
K^a(z_1) K^b(z_2) &=& \frac{\delta^{ab}}{z_{12}^2} + 
\frac{i\epsilon^{abc}}{z_{12}} J^c(z_2) + \cdots  \\
J^a(z_1) K^b(z_2) &=& \frac{i\epsilon^{abc}}{z_{12}} K^c(z_2) + \cdots 
\nonumber \end{eqnarray}
However, the fermionic components are not the same as in an $SO(4)$
SWZW theory.

\section{Conclusion}
In this paper we have  found new interesting features in $SU(N)$
 level $k = 0$ WZNW model and $SU(N)$ level  $k = N$ supersymmetric 
WZNW models using as toy models the $N=2$ cases.
 Besides logarithmic operators, there are new additional conserved
 currents generating new hidden symmetries. In the theories we have 
discussed, there are no operators with negative conformal dimensions,
 and so, in contrast to the majority of theories with logarithmic 
operators, these theories are not non-unitary. The further
investigation of these symmetries is of  great interest.

\section{Acknowledgements}
J.-S. Caux acknowledges support from NSERC Canada, and from the 
Rhodes Trust. A. Lewis acknowledges support from PPARC.
We would like to thank N. Mavromatos for interesting discussions.

\end{document}